\documentclass[useAMS,usenatbib]{mn2e}

\usepackage[dvips]{graphicx}
\usepackage[english]{babel}

\usepackage{graphicx}
\usepackage{color}

\title[Models of cuspy triaxial stellar systems. I. Stability and chaoticity]
{Models of cuspy triaxial stellar systems. I. Stability and chaoticity}
\author[A. F. Zorzi and J. C. Muzzio]{A. F. Zorzi$^{1}$\thanks{E-mail: 
azorzi@fceia.unr.edu.ar (AFZ); 
jcmuzzio@fcaglp.unlp.edu.ar (JCM)} and J. C. Muzzio$^{2}$\\
$^{1}$Instituto de F\'{i}sica de Rosario (CONICET--UNR) and Facultad de 
Ciencias Exactas, Ingenier\'{i}a y Agrimensura,\\
Universidad Nacional de Rosario, Rosario, Argentina\\
$^{2}$Instituto de Astrof\'{i}sica de La Plata (CONICET La Plata--UNLP) 
and Facultad de Ciencias Astron\'omicas y Geof\'{i}sicas, \\
Universidad Nacional de La Plata, La Plata, Argentina\\}
\begin{document}

\date{in original form 2012 January 10}

\pagerange{\pageref{firstpage}--\pageref{lastpage}} \pubyear{2012}

\maketitle

\label{firstpage}

\begin{abstract}
We used the N--body code of \citet{HO1992} to build a dozen cuspy 
($\gamma \simeq 1$) triaxial models of stellar systems through dissipationless 
collapses of initially spherical distributions of $10^6$ particles. We chose 
four sets of initial conditions that resulted in models morphologically resembling 
E2, E3, E4 and E5 galaxies, respectively. Within each set, three different seed 
numbers were selected for the random number generator used to create the initial 
conditions, so that the three models of each set are statistically
equivalent. We checked the stability of our models using the values of 
their central densities and of their moments of inertia, which turned out 
to be very constant indeed. The changes of those values were all less than 3 per cent 
over one Hubble time and, moreover, we show that the most likely cause of those 
changes are relaxation effects in the numerical code. We computed the six Lyapunov 
exponents of nearly 5,000 orbits in each model in order to recognize regular,
partially and fully chaotic orbits. All the models turned out to be highly chaotic, 
with less than 25 per cent of their orbits being regular. We conclude that it is 
quite possible to obtain cuspy triaxial stellar models that contain large 
fractions of chaotic orbits and are highly stable. The difficulty to build 
such models with the method of \citet{S1979} should be
attributed to the method itself and not to physical causes.

\end{abstract}

\begin{keywords}
Galaxies: elliptical and lenticular, cD -- Galaxies: kinematics and dynamics -- 
methods: numerical -- Physical data and processes: chaos.
\end{keywords}

\section{Introduction}

Self-consistent models of spherical and disklike stellar systems are relatively 
simple to build using standard textbook methods \citep{BT2008}, but special 
techniques are necessary to obtain triaxial models adequate to represent 
elliptical galaxies. One of the most popular of those techniques is due 
to \citet{S1979}: one chooses a reasonable mass distribution for the system, 
obtains the corresponding potential and computes a large library of orbits in that potential
selecting suitable initial conditions; weights are then assigned, according to the time spent
on each orbit and in every region of space, and used to set up a system of linear equations linking
the mass density in each region to the fractions of the different types of orbits, which can be
finally obtained solving the system. A different method, which we will refer to as the N--body
method, was due to \citet{SS1987}: adopting an initial distribution of mass points, one integrates
their equations of motion with an N--body code until an equilibrium configuration is reached;
the potential is then fixed and fitted with an adequate smooth approximation that allows the
computation of orbits using the positions and velocities of the bodies as initial conditions.
The initial distribution of bodies can be chosen in different ways to obtain a triaxial system:
slow deformation of a spherical system \citep{HMSH2001}, cold collapse of a spherical system
\citep*{VKS2002,MCW2005}, merging of stellar systems \citep*{JNB2005}, and so on.
Nevertheless, with the N--body method it is not possible to have the degree
of control over the final configuration that one has with Schwarzschild's method. In the end,
both methods yield the same result (i.e., a self-consistent stellar system and an analysis of its
orbital content), but the method of \citet{S1979} starts with a chosen mass distribution, obtains
orbits and finds which fractions of those orbits are needed to have a self-consistent system with
that mass distribution, while the N--body method begins obtaining a self-consistent system and then
performs the orbital analysis.

Chaotic motions are frequent in stellar systems \citep{C2004}, and \citet{S1993} realized that 
triaxial systems should include chaotic orbits but, when he introduced them
in his models, the models evolved on time scales of the order of a Hubble time and were not truly stable.
This problem was aggravated after it became clear that cuspy models (where near the centre the density,
$\rho(r)$, is proportional to $r^{-\gamma}$, where $r$ is the radius and $1 \leq \gamma \leq 2$) were needed
for elliptical galaxies \citep{MF1996}. Nevertheless, it is perfectly possible to obtain stable triaxial
models, even cuspy ones, that contain large fractions of chaotic orbits using the N--body method
(\citealt{VKS2002}; \citealt{MCW2005}; \citealt{AMNZ2007}; \citealt*{MNZ2009}), so that several of those authors 
argued that the difficulty to build such systems with the method of \citet{S1979} was due 
to that method itself and not to a physical cause.

Clearly, chaotic orbits are difficult to deal with when using the method of \citet{S1993} because
they cover different regions at different times. Worse still, in cuspy triaxial potentials chaotic
orbits tend to be extremely sticky \citep{SK2000}, and sticky orbits behave like regular orbits for
long periods of time and chaotically at other intervals. Let us assume, as an example, that chaotic
orbits occupy part of the time an elongated region of space and another part of the time a nearly
spherical region. Therefore, the orbits that occupy mostly an elongated region during the integration
time used to prepare the library of orbits will have large weights in that region and low weights
outside it; conversely, the orbits that occupy mostly the nearly spherical region during the same
integration time will have a more or less even distribution of weight values over this region. Now,
when these weights are used to establish the fractions of orbits that make up the triaxial system,
the orbits of the former group will be strongly favoured and the equilibrium system yielded by the
method of \cite{S1979} will include many of them and few of the other group. If we then let the
system evolve, the former orbits will tend to fill in a more spherical region as time goes by and,
conversely, the latter orbits will adopt a flatter distribution. But, since the model included
less orbits that originally had a more spherical distribution, the model will become rounder, as
shown by the Table 6 of \citet{S1993} for most of his models. The use of longer integration times,
as \citet{CLMV2007} advocated, does not guarantee the success of the method either, because the
average behaviour of a chaotic orbit over a certain interval, no matter how long, does not
necessarily coincide with its behaviour over a different interval, or even a subinterval, of the
integration time. To make matters worse, the number of orbits typically computed for Schwarzschild's
method is not very large, from several hundreds in the works of Schwarzschild himself up to about
$10,000$ in more recent work like that of \citet{CLMV2007}, so that they do not provide a strong
enough statistical basis.

On the other hand, the N--body method uses large numbers of bodies (of the order of $10^6$ in our
own more recent works) guaranteeing good statistics and the model is created by the evolution of
the system itself, that is, the dynamics take care of favoring certain orbits at the expense of
others so as to reach a self-consistent state. Is that state a stable one? If all the orbits are
regular it certainly is because, once reached that state, the orbits will continue filling in the
same regions of space for ever, in other words, we will have a static equilibrium. If the model
includes chaotic orbits, however, it is conceivable that, once in equilibrium, when part of 
the chaotic orbits evolves to occupy a rounder space region, another part of the chaotic orbits 
which had a rounder distribution evolves towards a more elongated one and fill in the phase 
space regions left vacant by the former, that is, we might have a dynamic rather than static 
equilibrium. While all this is concievable there is no guarantee that it will actually happen, 
but the past investigations that resulted in stable models that contained high fractions of 
chaotic orbits undoubtedly support this view.

Therefore, we want to investigate whether it is possible to obtain self-consistent models of cuspy
triaxial sytems that are stable. That is, we will deal with the type of model that is most likely
to contain high fractions of chaotic orbits and most difficult to obtain with Schwarzschild's
method. We have already obtained such models with the N--body method \citep{MNZ2009} but, in order
to compensate for the softening needed by the N--body code of L.A. Aguilar \citep{W1983, AM1990} we
had to introduce an additional potential. Here we will use the code of L. Hernquist \citep{HO1992}
that does not need softening and that uses an expansion of the potential based on the model of
\citet{H1990} that is particularly adequate for cuspy models.

The next section gives a description of how we obtained our models and their main properties.
Section 3 presents our results on the stability and chaoticity of the models and Section 4
summarizes our conclusions.

\section{Models of cuspy triaxial stellar systems}

\subsection{Model building}

We built our models following the recipe of \citet{AM1990}, just as we have done in our
previous investigations \citep[see][for example]{MCW2005, AMNZ2007, MNZ2009}: we randomly 
created a spherical distribution of $10^6$ particles with a density distribution 
inversely proportional to the distance to the centre and a Gaussian velocity distribution, 
and we let it collapse following the evolution with the code of \citet{HO1992}; 
due to the radial orbit instability, the result is a triaxial system. 
The gravitational constant, $G$, the radius of the sphere and the total
mass are all set equal to $1$ and the collapse time (the one needed to reach the maximum
potential energy) turns out to be also very close to unity. We followed the initial collapse
for 7 time units (t.u., hereafter) and, then, we eliminated the particles with positive
energy, we determined the principal axes of the inertia tensor of the $80$ per cent most tightly
bound particles and we rotated the system so that its major, intermediate and minor axes coincide,
respectively, with the x, y and z axes of coordinates. Subsequently, we let the system evolve for
another 300 crossing times ($T_{cr}$, hereafter, see Table 1 below),
eliminated the less bound particles that had not
yet reached equilibrium (less than $2$ or $3$ per cent of the particles in all cases) and aligned it again
with the system of coordinates. We then performed a last run of $600$ $T_{cr}$ to, first, obtain
the final model after the initial $100$ $T_{cr}$ and, second, check its stability over the final
$500$ $T_{cr}$. We must recall that we are only interested in obtaining models morphologically similar
to elliptical galaxies and not in creating them through a realistic process. Besides, we
want to make sure that our models have reached the equilibrium state before testing them
for possible evolutionary effects. As we will show later, one Hubble time is of the order
of $200$ $T_{cr}$ for our models, so that the integration times used here are indeed much larger
than that time and, as indicated, do not correspond to a realistic process of galaxy
formation.

After several trials, we chose an integration step of $0.0025 T_{cr}$ which yielded
integration errors in the energies of the individual bodies negligibly small compared to the changes
in those energies due to the relaxation effects of the code \citep{HB1990}. With that
choice, the total energy is conserved within about $0.1$ per cent during most of the evolutions,
except during the initial collapse when it is conserved only within about $1$ or $2$ per cent.

\citet*{KEV2008} investigated the approximation of N--body realizations of models
of \citet{D1993} for $0 \leq \gamma \leq 1$ with a generalization of the method of \citet{HO1992}
and they found that the choice of the radial basis functions seriously affects the results of
the fractions of chaotic orbits or the distribution of the Lyapunov characteristic exponents.
The model of \citet{H1990} corresponds to the $\gamma = 1$ case of the models of \citet{D1993}
so that the method of \citet{HO1992}, based on the former, should be expected to yield good results
for models with $\gamma = 1$ and, in fact, that is one conclusion of the work of \citet{KEV2008}.
Therefore, it is important that we stick to models with $\gamma = 1$, not only because they are
fairly cuspy, but also because the N--body method we are using works best for such models. Notice
that even our initial distribution (an sphere with density proportional to the inverse of the
radius) obeys that $\gamma$ value.

The code of \citet{HO1992} allows one to choose the number of terms in the angular and radial expansions
($l_{max}$ and $n_{max}$, respectively), so that we performed tests with different numbers of terms in
the ranges $2 \leq l_{max} \leq 6$ and $4 \leq n_{max} \leq 10 $ \citep{ZM2009}. For every
($l_{max}$, $n_{max}$) pair we run three simulations that differed only in the seed number used
to randomly generate the initial positions and velocities, i.e., the three were statistically
equivalent. The pairs $(3,6)$, $(3.8)$, $(3,10)$, $(4,7)$, $(4,10)$, $(5,8)$ and $(6,8)$ yielded
models that were only moderately cuspy, with $\gamma \simeq 0.6 - 0.9$, so that they were discarded.
It might seem odd that, despite the use of a basis of radial functions derived from the Hernquist
model, which has $\gamma = 1$, the obtention of collapse models with such cuspiness is not
guaranteed but, although the zero order radial function is cuspy with $\gamma=1$, higher order
terms are not and in all likelihood their contribution flattens the cusp. Most of the remaining
($l_{max}$, $n_{max}$) pairs might have been acceptable, as the reasons to prefer one to another were not
as compelling as the low $\gamma$ values that made us reject those mentioned before. Pairs $(2,8)$
and $(4,5)$ yielded $\gamma$ values that departed moderately (less than 0.1) from $1$, and the results
from their different statistical realizations exhibited larger dispersions than other pairs. Of the three
last pairs, all with $l_{max}=4$, the one with $n_{max}=6$ showed the most consistent results among the
different statistical realizations, so that we finally decided to use those numbers of terms in our expansions.

\subsection{The models}

Following \citet{AM1990} we took the square roots of the mean square values of coordinates
of the $80$ per cent most tightly bound particles as the semiaxes of the system; the major, intermediate
and minor axes will be dubbed $a$, $b$ and $c$, respectively, hereafter. We built several models
adopting different values of the dispersion for the velocity distribution, obtaining less elongated
models for larger dispersion values. Finally, we adopted four dispersion values that resulted
in models with $c/a$ values close to $0.8$, $0.7$, $0.6$ and $0.5$, respectively (i.e., corresponding to
elliptical galaxies between Hubble types E2 and E5). The velocity dispersion for the E5 model was
essentially zero, so that more elongated models could not be obtained. A similar result had been
obtained in our previous work \citep{AMNZ2007}, where we could not obtain models more flattened than
E6 and we suggested that this might hint that mergers, rather than collapses, are needed to obtain
the most elongated ellipticals. For each selected value of the velocity dispersion, three different models
(dubbed $a$, $b$ and $c$) were obtained using three different seed numbers for the random number generator
when creating the initial distribution of particles. In brief, we prepared a grand total of 12 models,
divided in 4 groups; the models in each group resemble E2, E3, E4 and E5 galaxies, respectively,
and the three models within each group differ from each other only from a microscopic point of view,
being essentially identical from the point of view of their macroscopic properties. As in our previous
work \citep{M2006, AMNZ2007, MNZ2009}, several models display figure rotation around the minor axis,
i.e., they rotate very slowly even though their total angular momentum is zero.

\begin{table*}
 \centering
 \begin{minipage}{140mm}
 \caption{Properties of our models.}
 \label{tab1}
 \begin{tabular}{@{}lccccccc@{}}
  \hline
$Model$ & $Mass$ & $T_{cr}$ & $R_{e}$ & $\sigma_{0}$ & $T$ & $\gamma$ & $\omega$ \\ 
\hline 
E2a  & 0.990 & 0.721 & 0.250 & 0.894 & 0.73 & 1.052 $\pm$ 0.021 & 0.0003 $\pm$ 0.0001 \\
E2b  & 0.990 & 0.723 & 0.253 & 0.897 & 0.78 & 0.997 $\pm$ 0.021 & 0.0012 $\pm$ 0.0004 \\ 
E2c  & 0.990 & 0.721 & 0.250 & 0.902 & 0.70 & 1.006 $\pm$ 0.021 & 0.0016 $\pm$ 0.0004 \\ 
\\
E3a  & 0.979 & 0.659 & 0.220 & 0.934 & 0.65 & 1.056 $\pm$ 0.021 & -0.0004 $\pm$ 0.0001 \\ 
E3b  & 0.979 & 0.659 & 0.221 & 0.933 & 0.68 & 1.062 $\pm$ 0.021 & 0.0002 $\pm$ 0.0001 \\
E3c  & 0.978 & 0.657 & 0.219 & 0.929 & 0.66 & 0.987 $\pm$ 0.021 & 0.0010 $\pm$ 0.0001 \\
\\
E4a  & 0.911 & 0.526 & 0.182 & 0.957 & 0.62 & 1.053 $\pm$ 0.019 & 0.0001 $\pm$ 0.0003 \\
E4b  & 0.905 & 0.518 & 0.180 & 0.974 & 0.63 & 1.027 $\pm$ 0.021 & -0.0002 $\pm$ 0.0001 \\
E4c  & 0.908 & 0.521 & 0.180 & 0.963 & 0.62 & 1.073 $\pm$ 0.022 & -0.0002 $\pm$ 0.0002 \\
\\
E5a  & 0.906 & 0.463 & 0.159 & 0.997 & 0.46 & 0.985 $\pm$ 0.022 & 0.0020 $\pm$ 0.0007\\
E5b  & 0.908 & 0.466 & 0.160 & 0.986 & 0.48 & 1.000 $\pm$ 0.019 & 0.0059 $\pm$ 0.0004 \\
E5c  & 0.907 & 0.465 & 0.158 & 0.970 & 0.46 & 1.017 $\pm$ 0.020 & 0.0164 $\pm$ 0.0003 \\
\hline
\end{tabular}
\end{minipage}
\end{table*}

Table 1 summarizes the global properties of our models: mass, crossing time, effective radius,
central radial velocity dispersion, triaxiality, $\gamma$ and angular velocity of figure rotation. The effective
radius was obtained from the $(x, z)$ projection and, accordingly, the central radial velocity dispersion
was computed from the y components of the velocities of the $10,000$ particles closer to the centre on
that projection. Triaxiality was evaluated from the semiaxes obtained from the $80$ per cent most tightly bound
particles as $T = (a^2 -b^2)/(a^2-c^2)$. $\gamma$ was obtained as the slope of the
$log \rho(r)$ vs. $log (r)$ line for the
innermost $10,000$ particles binned in $100$ particle bins. The angular velocity was computed from the
angles formed by the major axis with its original position at different times of the final $500$ $T_{cr}$
integration. The Table clearly shows that, for a given model, the different realizations obtained
changing the seed number in the random number generator have essentially the same global properties.
The only exception is the angular velocity which displays significant differences, particularly among models
E5a, b and c. Nevertheless, rotation is very slow in all cases and, except for the E5c case, lower than the
0.00975 value of the investigation by \citet{M2006}, which revealed only very small differences between the
fractions of chaotic orbits in the rotating and non-rotating system. The present E4 and E5 models are
much more triaxial than the corresponding non-cuspy models of \citet{AMNZ2007} ($0.98$ and $0.81$, 
respectively), and the present model E4 is much more triaxial than the cuspy model E4c of \citet{MNZ2009} 
($0.91$).

As in our previous work \citep[see][for example]{AMNZ2007, MNZ2009}, we chose galaxies
NGC1379 and NGC4697 \citep{NCR2005, FP1999},  whose mass-to-light ratio gradients are zero,
to obtain some estimate of the equivalence between our units and those of real galaxies,
Comparing their observed values of $R_{e}$ ($2.5$ and $5.7 kpc$, respectively)
and $\sigma_{0}$ ($128$ and $180 \: km s^{-1}$, respectively) with those from our Table~\ref{tab1}, 
we conclude that values between about $10$ and $36 kpc$ can be used as our length unit and 
values between about $0.07$ and $0.20 Gy$ as our time unit. Then, the Hubble 
time can be estimated as between $66$ and $190 \:t.u.$, and we will adopt a value of $100 \:t.u.$,
hereafter.
\begin{table}
\caption{Major semiaxes and axial ratios of our models}
\label{tab2}
\begin{tabular}{@{}lcccccc@{}}
\hline
$Model$ & $Property$ & $20\%$ & $40\%$ & $60\%$ & $80\%$ & $100\%$ \\ 
\hline 
$ $ & a   & 0.068 & 0.117 & 0.172 & 0.300 & 0.578 \\ 
E2a & b/a & 0.754 & 0.789 & 0.837 & 0.877 & 0.930 \\ 
$ $ & c/a & 0.596 & 0.684 & 0.768 & 0.826 & 0.909 \\
\\
$ $ & a   & 0.068 & 0.118 & 0.175 & 0.297 & 0.574 \\ 
E2b & b/a & 0.742 & 0.777 & 0.820 & 0.870 & 0.925 \\ 
$ $ & c/a & 0.602 & 0.687 & 0.763 & 0.829 & 0.906 \\
\\
$ $ & a   & 0.067 & 0.116 & 0.173 & 0.294 & 0.578 \\ 
E2c & b/a & 0.767 & 0.801 & 0.845 & 0.882 & 0.937 \\ 
$ $ & c/a & 0.605 & 0.691 & 0.769 & 0.826 & 0.913 \\
  \hline
$ $ & a   & 0.062 & 0.115 & 0.164 & 0.294 & 1.097 \\ 
E3a & b/a & 0.753 & 0.723 & 0.781 & 0.814 & 0.852 \\ 
$ $ & c/a & 0.597 & 0.564 & 0.642 & 0.694 & 0.810 \\
\\
$ $ & a   & 0.065 & 0.117 & 0.165 & 0.294 & 1.139 \\ 
E3b & b/a & 0.710 & 0.701 & 0.761 & 0.802 & 0.840 \\ 
$ $ & c/a & 0.554 & 0.555 & 0.634 & 0.692 & 0.803 \\
\\
$ $ & a   & 0.065 & 0.115 & 0.165 & 0.294 & 1.144 \\ 
E3c & b/a & 0.714 & 0.715 & 0.776 & 0.808 & 0.852 \\ 
$ $ & c/a & 0.559 & 0.559 & 0.635 & 0.689 & 0.807 \\
\hline
$ $ & a   & 0.056 & 0.104 & 0.152 & 0.233 & 1.146 \\ 
E4a & b/a & 0.731 & 0.697 & 0.735 & 0.769 & 0.775 \\ 
$ $ & c/a & 0.581 & 0.516 & 0.546 & 0.581 & 0.674 \\
\\
$ $ & a   & 0.056 & 0.102 & 0.150 & 0.231 & 1.160 \\ 
E4b & b/a & 0.728 & 0.698 & 0.733 & 0.764 & 0.792 \\ 
$ $ & c/a & 0.575 & 0.517 & 0.549 & 0.583 & 0.706 \\
\\
$ $ & a   & 0.055 & 0.103 & 0.152 & 0.232 & 1.180 \\ 
E4c & b/a & 0.749 & 0.706 & 0.737 & 0.765 & 0.797 \\ 
$ $ & c/a & 0.588 & 0.515 & 0.539 & 0.575 & 0.708 \\
\hline
$ $ & a   & 0.052 & 0.095 & 0.148 & 0.216 & 0.482 \\ 
E5a & b/a & 0.824 & 0.809 & 0.815 & 0.814 & 0.893 \\ 
$ $ & c/a & 0.557 & 0.508 & 0.504 & 0.515 & 0.563 \\
\\
$ $ & a   & 0.051 & 0.095 & 0.147 & 0.222 & 0.479 \\ 
E5b & b/a & 0.846 & 0.810 & 0.802 & 0.803 & 0.879 \\
$ $ & c/a & 0.596 & 0.509 & 0.506 & 0.508 & 0.555 \\
\\
$ $ & a   & 0.051 & 0.094 & 0.145 & 0.222 & 0.496 \\ 
E5c & b/a & 0.847 & 0.825 & 0.810 & 0.810 & 0.883 \\ 
$ $ & c/a & 0.583 & 0.512 & 0.500 & 0.506 & 0.569  \\
\hline
\end{tabular}
\end{table}

\begin{figure}
\vspace{20pt}
\includegraphics[width=84mm]{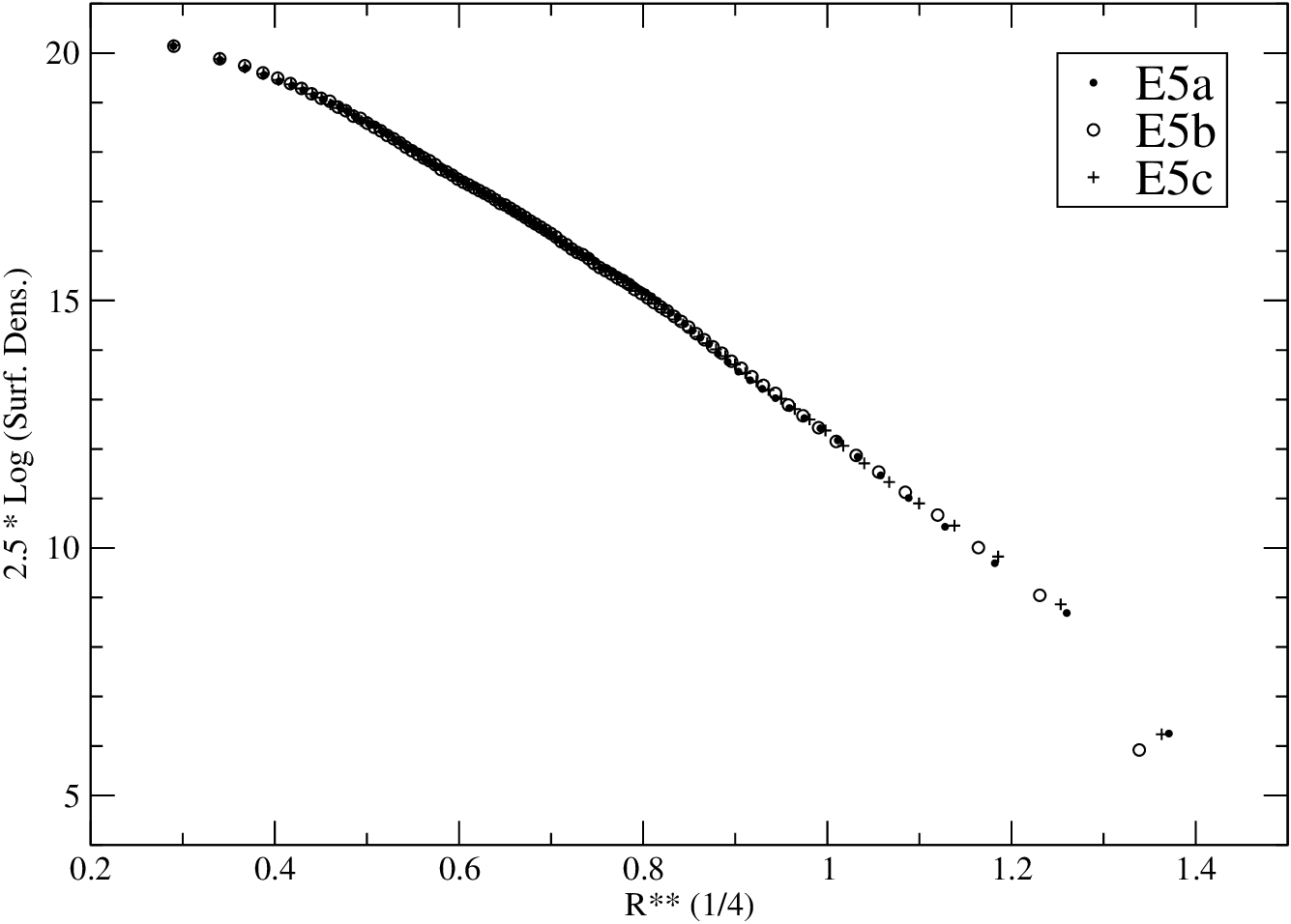}
\caption{Surface density versus $R^{1/4}$ plot for the E5 models}
	\label{fig:leyvaucE5}
\end{figure}

We obtained the projected distributions of the particles on the $(xz)$
plane and, following \citet{AM1990}, we adopted fixed axial ratios equal
to those of the $80$ per cent most tightly bound particles for each model (see
below), and computed the surface density of particles in elliptical shells 
containing $10,000$ particles per bin. Just as \citet{AM1990} had found
for their models, ours follow very closely a de Vaucouleurs law and, again,
differences among the different random realization of the same model are
negligibly small. As an example, Figure \ref{fig:leyvaucE5} shows the results
for our models E5.

\begin{figure}
\vspace{20pt}
\includegraphics[width=84mm]{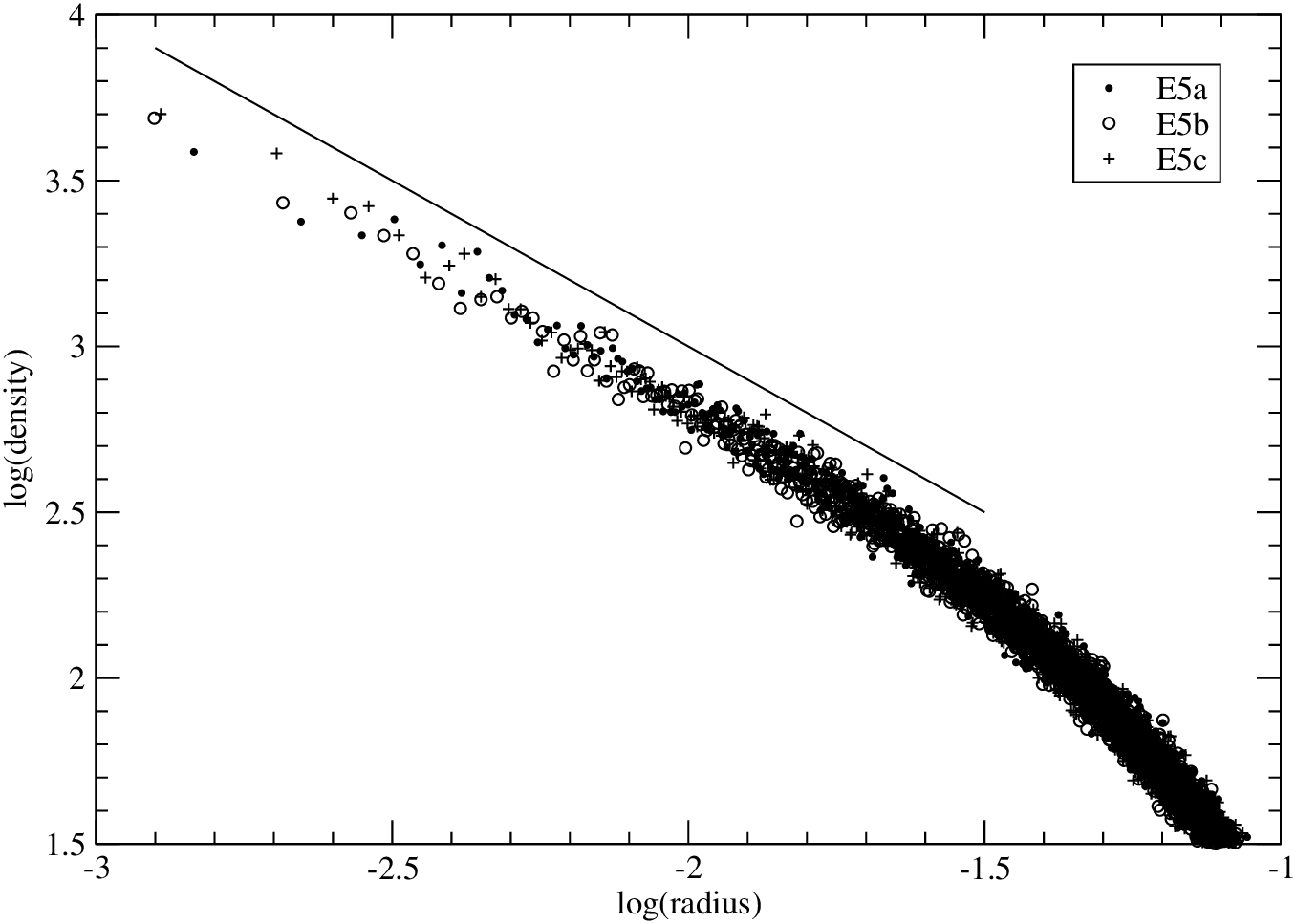}
\caption{Density versus radius plot for the E5 models}
	\label{fig:densvsradE5}
\end{figure}

\begin{figure}
\vspace{20pt}
\includegraphics[width=84mm]{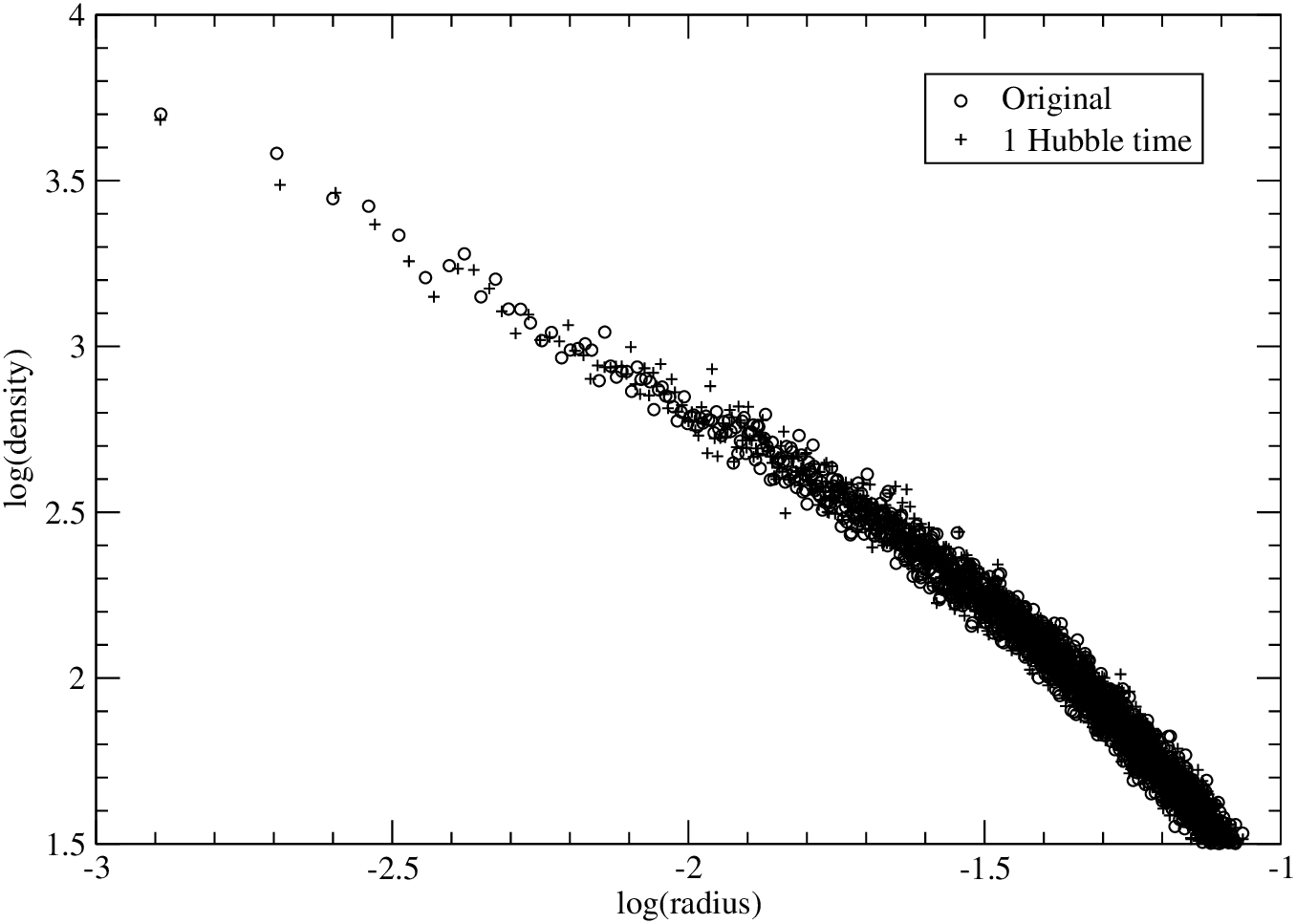}
\caption{Density versus radius plot for the E5c model at different times.}
	\label{fig:densvsradE5c}
\end{figure}

Figure~\ref{fig:densvsradE5} shows, as an example, the logarithm of the density
versus the logarithm of the radius for the central part of the E5 models; different
symbols correspond to models with different seed numbers and a straight line with
slope $\gamma = 1$ is also shown for comparison. In this case the density was computed in
spherical shells containing 100 particles each, so that the Poisson relative error
of each point on the graph is about $0.043$.
Besides, Figure~\ref{fig:densvsradE5c} presents, for the innermost region of model E5c,
the logarithm of the density versus the logarithm of the radius for both the adopted model 
(circles) as for that same model evolved $100 \:t.u.$, or about a Hubble time, 
(plus signs) and we note that the slope of the central cusp is very well conserved. 
Similar results were obtained for the other models.

Table~\ref{tab2} gives for each model the values of the major semiaxis and of the axial ratios 
for the $20$, $40$, ...and  $100$ per cent most tightly bound particles. Again, the results
obtained changing the randomly generated inital conditions for a given model are essentially
the same. As in our previous models, there is a general trend towards larger axial ratios at
larger distances from the centre, but the trend is weaker for the more flatened models and
breaks closer to the centre, probably due to the influence of the cusp.

\section{Results and analysis}
\subsection{Stability}
We used the results of the $500$ $T_{cr}$ long final evolution to check the stability of 
the central density and the semiaxes of the stellar systems. For this purpose, we computed at
$100$ $T_{cr}$ intervals the central density from the $10,000$ particles closer to
the centre of each system, and the moments of inertia of the $80$ per cent most tightly bound particles.
These quantities changed almost linearly with time, so that we obtained their variations over a
Hubble time from the corresponding best fitting straight lines and the results are presented in
Table~\ref{tab3}.

\begin{table*}
\centering
 \begin{minipage}{140mm}
\caption{Percentage variations over one Hubble time.} 
\label{tab3}
\begin{tabular}{@{}lcccc@{}}
\hline
$Model$ & $Dens. (\%)$& $X mom. in. (\%)$& $Y mom. in. (\%)$ & $Z mom. in. (\%)$ \\ 
\hline 
 E2a & -0.61 $\pm$ 0.29 & -0.63 $\pm$ 0.09 & 0.64 $\pm$ 0.13 & 0.20 $\pm$ 0.09 \\ 
 E2b & -0.36 $\pm$ 0.43 & -0.48 $\pm$ 0.07 & 0.51 $\pm$ 0.08 & 0.41 $\pm$ 0.11 \\ 
 E2c & -0.83 $\pm$ 0.59 & -0.45 $\pm$ 0.10 & 0.49 $\pm$ 0.10 & 0.40 $\pm$ 0.13 \\ 
 \\
 E3a & -0.76 $\pm$ 0.53 & -1.01 $\pm$ 0.05 & 1.36 $\pm$ 0.11 & 0.78 $\pm$ 0.10 \\ 
 E3b & -0.62 $\pm$ 0.15 & -1.08 $\pm$ 0.07 & 1.44 $\pm$ 0.09 & 0.86 $\pm$ 0.15 \\ 
 E3c & -0.56 $\pm$ 0.18 & -1.04 $\pm$ 0.11 & 1.57 $\pm$ 0.16 & 0.87 $\pm$ 0.11 \\
\\ 
 E4a & -1.05 $\pm$ 0.65 & -1.12 $\pm$ 0.06 & 1.16 $\pm$ 0.04 & 1.68 $\pm$ 0.12 \\
 E4b & -0.86 $\pm$ 0.86 & -1.21 $\pm$ 0.09 & 1.19 $\pm$ 0.08 & 1.65 $\pm$ 0.17 \\
 E4c & -0.73 $\pm$ 0.56 & -1.20 $\pm$ 0.10 & 1.36 $\pm$ 0.23 & 1.70 $\pm$ 0.18 \\
 \\  
 E5a & -1.07 $\pm$ 0.78 & -2.17 $\pm$ 0.11 & 2.12 $\pm$ 0.14 & 2.65 $\pm$ 0.21 \\
 E5b &  0.92 $\pm$ 0.83 & -1.96 $\pm$ 0.13 & 2.00 $\pm$ 0.08 & 1.95 $\pm$ 0.08 \\
 E5c & -1.23 $\pm$ 0.17 & -2.02 $\pm$ 0.09 & 1.73 $\pm$ 0.21 & 2.31 $\pm$ 0.06 \\ 
\hline
\end{tabular} 
\end{minipage}
\end{table*}

Although most of the variations of the central density are not significant at
the $3\sigma$ level it is suggestive that, except for that of model E5b (also
not significant), they are all negative. Almost all the variations of the moments
of inertia are highly significant and indicate a general decrease of the major
axes as well as general increases of the intermediate and minor axes. But
the changes are very small indeed, all of them being smaller than $3$ per cent over
one Hubble time.

Interestingly, both the amounts and the senses of the changes are similar to those
found in our previous works \citep{AMNZ2007, MNZ2009}, where we attributed them mainly
to relaxation effects of the multipolar code \citep{HB1990}. Since in those works we
had used the N--body code of Aguilar and we are now using that of Hernquist and Ostriker,
we decided to perform the same checks we had made before using the models E5, i.e., those that
exhibit the largest changes. The first check was to eliminate self-consistency, letting the
systems evolve again for $500$ $T_{cr}$ but keeping the coefficients of the expansion of the
potential fixed at their initial values, and the results are shown in Table~\ref{tab4}.
The central density changes, although somewhat lower than the corresponding values of
Table~\ref{tab3}, are again not significant at the $3\sigma$ level, but the changes
of the moments of inertia are much lower and generally significant at the same level.
This is what could be expected from changes due to relaxation effects of the N--body
code, because they would be suppressed when turning off the self-consistency. Alternatively,
relaxation effects should increase when the number of bodies decreases, and that was
our second check. We took at random $10$ per cent of the particles of each one of the models E5,
increased their masses 10 times, and we run these new models self-consistently
first for $150$ $T_{cr}$ to let them relax, and then for another $500$ $T_{cr}$ to analyze their
stability. The corresponding changes are shown in Table~\ref{tab5} and, except for
those of the central density which remain not significant, they are substantially larger
than the equivalente ones of Table~\ref{tab3}. Thus, we may conclude that even the small
variations shown in the latter Table, are mainly due to relaxation effects of the N--body code.
  
\begin{table*}
\centering
\begin{minipage}{140mm}          
\caption{Percentage variations over one Hubble time, for models E5 with constant coefficients.} 
\label{tab4}
\begin{tabular}{@{}lcccc@{}}
\hline
$Mod$ & $Dens. (\%)$& $X mom. in. (\%)$& $Y mom. in. (\%)$ & $Z mom. in. (\%)$ \\ 
\hline 
 E5a &  +0.02 $\pm$ 0.74 &-0.29 $\pm$ +0.09 & -0.08 $\pm$ 0.10 &  +0.41 $\pm$ 0.07 \\
 E5b &  +0.46 $\pm$ 0.80 &-0.33 $\pm$ +0.06 & -0.01 $\pm$ 0.10 &  +0.38 $\pm$ 0.13 \\
 E5c &  -0.96 $\pm$ 0.44 &-0.26 $\pm$ +0.05 & -0.15 $\pm$ 0.12 &  +0.46 $\pm$ 0.10 \\  
\hline
\end{tabular}
\end{minipage}
\end{table*}

\begin{table*}
\centering
\begin{minipage}{140mm}
\caption{Percentage variations over one Hubble time, for models E5 with ten times less bodies.} 
\label{tab5}
\begin{tabular}{@{}lcccc@{}}
\hline
$Mod$ & $Dens. (\%)$& $X mom. in. (\%)$& $Y mom. in. (\%)$ & $Z mom. in. (\%)$ \\ 
\hline 
 E5a &  +1.18 $\pm$ 0.83 &-4.98 $\pm$ 0.50 &  +5.70 $\pm$ 0.54 &  +5.47 $\pm$ 0.53 \\
 E5b &  +0.07 $\pm$ 1.11 &-4.75 $\pm$ 0.53 &  +5.64 $\pm$ 0.45 &  +6.38 $\pm$ 0.41 \\
 E5c &  +0.68 $\pm$ 0.87 &-5.32 $\pm$ 0.46 &  +5.21 $\pm$ 0.28 &  +6.46 $\pm$ 0.76 \\
\hline
\end{tabular} 
\end{minipage}
\end{table*}

\subsection{Regular, partially and fully chaotic orbits}
We randomly selected between $4,500$ and $5,000$ particles from each model and 
adopted their positions and velocities as the initial values to obtain the orbits 
and investigate their chaoticity. The potentials were fixed,
keeping constant the coefficients of their
expansions at their final values, and the integrations were carried out in fixed
coordinate systems,  in those cases where the rotation velocity was not significant
(i.e., less than three times the mean square error), or in systems rotating with
the corresponding velocity in those cases where it was significant. 
We proceeded in this way to be consistent with the models obtained, but even the significant
velocities are so small that their effect is in all likelihood negligible \citep{M2006}.

Since the potentials of our systems were fixed, all our orbits obey the energy integral
(or the Jacobi integral in the case of systems with significant rotational velocity).
Therefore, regular orbits have to obey two additional isolating integrals, but we can have
two kinds of chaotic orbits: Partially chaotic orbits obey only one additional integral
besides energy, and fully chaotic orbits have no isolating integrals other than energy.
We have shown before \citep[see][for example]{MCW2005, AMNZ2007, MNZ2009} that partially 
and fully chaotic orbits have different distributions, so that it is important to 
separate them in orbital structure studies. A simple, albeit computing time demanding, 
way of classifying regular, partially and fully chaotic orbits is through the use of the six
Lyapunov exponents. Since phase space volume is conserved, the exponents come
in three pairs  of the same absolute value and opposite sign. Due to energy conservation, one
of those  pairs is always zero in our case, and each additional isolating integral makes zero 
another pair. Thus, regular orbits have all their Lyapunov exponents equal to zero, 
partially chaotic orbits have one non-zero pair and fully chaotic orbits have two.

The numerical equivalent of the Lyapunov exponents (which demand to integrate the orbit
over an infinite time interval) are the finite time Lyapunov
characteristic numbers (hereafter FT-LCNs) and, as in our previous works, we computed them
using the LIAMAG subroutine \citep{UP1988}, kindly provided by D. Pfenniger, and that we
adapted with the assistance of H.D. Navone to include the potential, accelerations and
variational equations corresponding to the method of Hernquist and Ostriker.
The LIAMAG uses double precision arithmetic and a Runge - Kutta - Fehlberg
integrator of order 7-8 with variable time steps. As in our previous works, we adopted the
integration and normalization intervals as $10,000 \:t.u.$ and $1 \:t.u.$, respectively, and
we requested a precision of $10^{-15}$ for the step size regulation, which resulted in an
energy conservation better than $2\times10^{-12}$ at the end of integration in all cases.
We will refer to the largest FT-LCN of a given orbit as $L_{max}$ and the second largest one
as $L_{int}$, hereafter. 

\begin{figure}
\vspace{20pt}
\includegraphics[width=84mm]{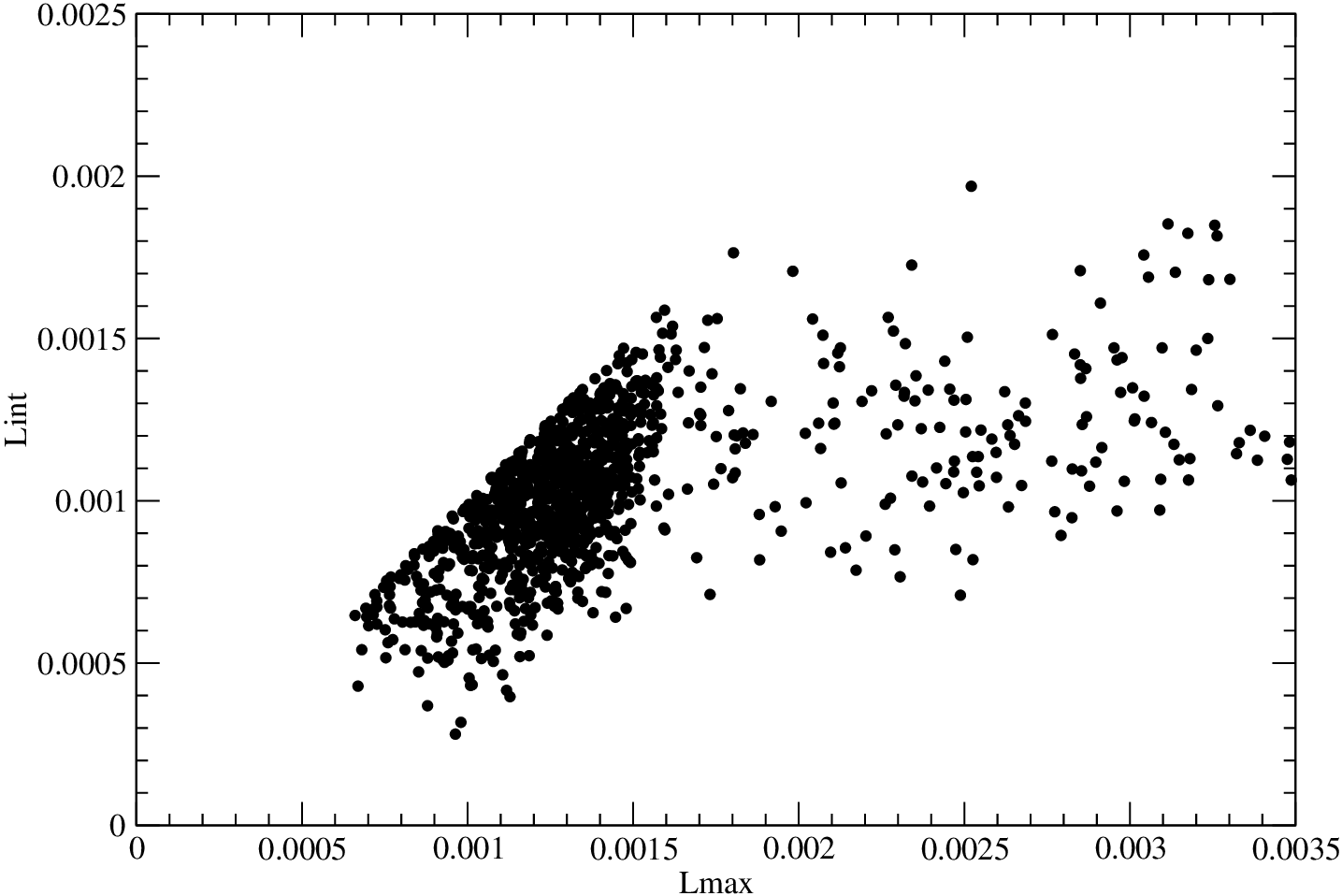}
\caption{$L_{max}$ versus $L_{int}$ for Model E5c. Only the region of low values,
   around the region corresponding to regular orbits is shown.}
	\label{fig:lya5abc}
\end{figure}

Since the FT-LCNs are obtained from numerical integrations over 
a finite time interval (i.e., $10,000 \:t.u.$ here), rather than the infinite 
one required to obtain Lyapunov exponents, they cannot reach zero value, but only a
limiting minimum value, $L_{lim}$. Applying the definition of
the Lyapunov exponents to equation (7) of \citet{CGS2003}, one can estimate
that value as $L_{lim} \approx lnT/T$, where T is the integration interval (Cincotta,
private communication), that is, a limiting value of about $0.00092 \: (t.u.)^{-1}$ for
our $T = 10,000 \:t.u.$ interval. This is an order of magnitude estimate only, however, and it is
better to derive a more accurate value from the FT-LCNs themselves using, for example,
plots of the low end of the $L_{int}$ versus $L_{max}$ distribution (i.e., the region around
that occupied by the representative points of the regular orbits). As an example we show
in Figure~\ref{fig:lya5abc} the corresponding plot for model E5c. The representative points
of the regular orbits are concentrated on the blob at the left and those of chaotic orbits
extend towards the right (the sharp envelope of the identity line is merely due to the fact that,
by definition, $L_{int} \leq L_{max}$). Clearly, any limit that attempts to separate regular
from chaotic orbits can have a statistical value only, because some regular orbits may have FT-LCNs
slightly larger than that limit, while some chaotic orbits may have somewhat lower FT-LCNs. With
that caveat, using plots equivalent to that of Figure~\ref{fig:lya5abc} for all our models we
adopted $L_{lim}=0.0018 \:(t.u.)^{-1}$.

Now, that $L_{lim}$ corresponds to a Lyapunov time of $556 \: t.u.$, which is equivalent to
about 5 or 6 Hubble times for our models. Is it reasonable to use such a low $L_{lim}$
to separate regular from chaotic orbits? Would it not be more sensible to adopt a
value of $0.0100  \: (t.u.)^{-1}$ which corresponds to a Lyapunov time of the same order
of the Hubble time? Those questions have been considered in our previous work
\citep[see][for example]{AMNZ2007, MNZ2009} and we will repeat here the same analysis done there.

First, we separated the orbits of each model into three groups: a) Those with
$Lmax < 0.0018 \:(t.u.)^{-1}$, i.e., those that are classified as regular for both choices
of $L_{lim}$ (REGREG, hereafter); b) Those with 
$0.0018 \:(t.u.)^{-1}~\leq ~ L_{max}~<~0.0100 \:(t.u.)^{-1}$,
i.e., those that are classified as regular  for $L_{lim}=0.0100 \:(t.u.)^{-1}$, but as
chaotic for $L_{lim}=0.0018 \:(t.u.)^{-1}$ (REGCHAO, hereafter);  c) Those with
$0.0100 \:(t.u.)^{-1}~\leq~L_{max}$, ie, those that are classified as chaotic for both
elections of $L_{lim}$ (hereafter CHAOCHAO). Then we considered, for each orbit,
11 $(x,~ y,~ z)$ orbital positions separated by intervals of $10 \:t.u.$, that is, over a
total interval of $100\: t.u.$, and, for
each model and for each type of orbit, we computed the mean square value of each coordinate.
Table ~\ref{tab6} gives the square roots of the ratios of the $y$ and $z$ mean square values
to the $x$ mean square value.

\begin{table}
\caption{Axial ratios of different classes of orbits for different choices of $L_{lim}$.}
\label{tab6} 
\begin{tabular}{@{}lcccc@{}}
\hline
$Ratio$ & $Mod$ & $REGREG$    &     $REGCHAO$     & $CHAOCHAO$ \\ 
\hline 
y/x & E2a & 1.030 $\pm$ 0.015 & 0.868 $\pm$ 0.013 & 0.865 $\pm$ 0.015 \\ 
$ $ & E2b & 1.023 $\pm$ 0.014 & 0.868 $\pm$ 0.013 & 0.875 $\pm$ 0.015 \\ 
$ $ & E2c & 1.003 $\pm$ 0.014 & 0.847 $\pm$ 0.014 & 0.856 $\pm$ 0.014 \\
 \\
$ $ & E3a & 0.816 $\pm$ 0.022 & 0.851 $\pm$ 0.018 & 0.856 $\pm$ 0.018 \\ 
$ $ & E3b & 0.815 $\pm$ 0.023 & 0.824 $\pm$ 0.019 & 0.829 $\pm$ 0.017 \\ 
$ $ & E3c & 0.773 $\pm$ 0.025 & 0.866 $\pm$ 0.018 & 0.877 $\pm$ 0.016 \\
\\
$ $ & E4a & 0.520 $\pm$ 0.040 & 0.745 $\pm$ 0.026 & 0.874 $\pm$ 0.019 \\
$ $ & E4b & 0.510 $\pm$ 0.044 & 0.788 $\pm$ 0.028 & 0.841 $\pm$ 0.019 \\
$ $ & E4c & 0.568 $\pm$ 0.039 & 0.765 $\pm$ 0.023 & 0.830 $\pm$ 0.019 \\
 \\
$ $ & E5a & 0.892 $\pm$ 0.028 & 0.877 $\pm$ 0.025 & 0.908 $\pm$ 0.018 \\
$ $ & E5b & 0.965 $\pm$ 0.026 & 0.833 $\pm$ 0.025 & 0.899 $\pm$ 0.018 \\
$ $ & E5c & 0.934 $\pm$ 0.029 & 0.759 $\pm$ 0.025 & 0.839 $\pm$ 0.016 \\ 
 \hline
z/x & E2a & 0.910 $\pm$ 0.016 & 0.915 $\pm$ 0.013 & 0.905 $\pm$ 0.015 \\ 
$ $ & E2b & 0.937 $\pm$ 0.015 & 0.903 $\pm$ 0.013 & 0.874 $\pm$ 0.015 \\ 
$ $ & E2c & 0.879 $\pm$ 0.015 & 0.928 $\pm$ 0.014 & 0.891 $\pm$ 0.014 \\
\\
$ $ & E3a & 0.676 $\pm$ 0.024 & 0.840 $\pm$ 0.018 & 0.832 $\pm$ 0.019 \\ 
$ $ & E3b & 0.712 $\pm$ 0.027 & 0.812 $\pm$ 0.018 & 0.771 $\pm$ 0.018 \\ 
$ $ & E3c & 0.668 $\pm$ 0.028 & 0.851 $\pm$ 0.018 & 0.823 $\pm$ 0.017 \\
 \\ 
$ $ & E4a & 0.424 $\pm$ 0.053 & 0.719 $\pm$ 0.028 & 0.760 $\pm$ 0.020 \\
$ $ & E4b & 0.353 $\pm$ 0.044 & 0.660 $\pm$ 0.028 & 0.767 $\pm$ 0.019 \\
$ $ & E4c & 0.401 $\pm$ 0.049 & 0.710 $\pm$ 0.026 & 0.769 $\pm$ 0.020 \\ 
 \\
$ $ & E5a & 0.309 $\pm$ 0.033 & 0.501 $\pm$ 0.031 & 0.759 $\pm$ 0.021 \\
$ $ & E5b & 0.361 $\pm$ 0.040 & 0.527 $\pm$ 0.035 & 0.699 $\pm$ 0.020 \\
$ $ & E5c & 0.319 $\pm$ 0.043 & 0.473 $\pm$ 0.032 & 0.699 $\pm$ 0.018 \\
\hline
\end{tabular}
\end{table}

The results in Table \ref{tab6} show that, as in our previous work,
most of the axial ratios of the REGCHAO orbits are significantly different
from those of the REGREG orbits and we may conclude that, despite their
low FT-LCN values implying Lyapunov times longer than the Hubble time,
orbits with $0.0018 \:(t.u.)^{-1} \leq L_{max} < 0.0100 \:(t.u.)^{-1}$ have a
spatial distribution different from that of regular orbits. In other words,
the time scale for the exponential divergence of orbits (measured by the
Lyapunov time) is not much relevant for the spatial distribution of those
orbits. Since we are here interested in that distribution, the sensible
approach is thus to adopt $L_{lim} = 0.0018 \:(t.u.)^{-1}$. 
Therefore, we classify orbits as regular  if $L_{max}~ <~ L_{lim}$, 
as partially chaotic if $L_{int}~ <~ L_{lim}~ \leq~ L_{max}$  and 
as fully chaotic if $L_{lim}~ \leq~ L_{int}$.

Table \ref{tab7} gives the percentages of regular, partially and fully chaotic
orbits in our models. As in our previous works, the statistical errors have been
estimated from the binomial distribution. The agreement among results of the a, b
and c cases of each model is excellent and within the $3\sigma$ level in all cases,
except for the regular and fully chaotic orbits of the E3c case where the differences
with the E3a and E3b cases are of the order of $5\sigma$. Clearly, all our models
harbor the lowest fractions of regular orbits in triaxial systems we know about,
lower even than the $29.05$ per cent found by \citet{MNZ2009} for their model E6c.
Interestingly, the fraction of regular orbits does not decrease monotonically
from type E2 through type E5, but falls first to rise again; a somewhat similar, but
less pronounced, trend was found in the E4 through E6 models of \citet{AMNZ2007}.
The decreasing fractions of partially chaotic orbits when going from type E2 through
E5 show the same trend found in those two previous works of ours.

Two further tests were performed to check the accuracy of our results.
On the one hand, it is interesting to check the influence that the use of a particular
N-body snapshot as a model might have. On the other hand, since we are using only about
0.5 per cent of the orbits in each model to investigate chaoticity, it might be worthwile
to see if the use of a different sample yields different results. We adopted the E4b model
for both tests. For the first one, we took a new model from the snapshot evolved $200 T_{cr}$
more than the one whose results are presented in Table \ref{tab7}, and we obtained fractions
of $14.42 \pm 0.52\%$, $11.95 \pm 0.48\%$ and $73.62 \pm 0.65\%$, respectively for the regular,
partially and fully chaotic orbits. The first two fractions coincide with those of the original
model at the $3\sigma$ level, but the last one differs by $3.4\sigma$. Not only are these
differences small, but they are probably at least partially due to the axial changes of the
models as they evolve described in Subsection 3.1. The $200 T_{cr}$ evolution changes the
triaxiality of the E4b model in the direction of that of the E5 models by about $1/10$ of the
triaxiality difference between the E4 and E5 models. Since the difference between the fractions
of fully chaotic orbits of those models is about $8\%$ we might crudely estimate that about
$0.8\%$ of the change from the original model to the one evolved $200 T_{cr}$ might be attributed
to the axial changes, in which case the remaining difference is of $2.5\sigma$ only. Nevertheless,
the agreement between the original and the $200 T_{cr}$ evolution results is somewhat poorer than
those among statistically different realizations of the same model, except for the already
mentioned differences for models E3. The second test, instead, yielded no significant differences.
The new sample of orbits from E4b model had fractions of $12.50 \pm 0.49\%$, $9.74 \pm 0.44\%$ and
$77.76 \pm 0.62\%$, respectively of regular, partially and full chaotic orbits, all of them well
within the $3\sigma$ level of differences from the original results.

\begin{table}
\caption{Percentages of regular and chaotic orbits in triaxial systems.}
\label{tab7}
\begin{tabular}{@{}lccc@{}}
\hline
$Mod$ & $Regular(\%)$ & $Part. \;Chaotic (\%)$ & $Fully \;Chaotic (\%)$ \\ 
\hline 
E2a &  22.48 $\pm$ 0.59   &  15.30 $\pm$ 0.51   &  62.22 $\pm$ 0.69   \\ 
E2b &  21.35 $\pm$ 0.58   &  15.52 $\pm$ 0.51   &  63.13 $\pm$ 0.69   \\ 
E2c &  22.24 $\pm$ 0.59   &  13.64 $\pm$ 0.49   &  64.12 $\pm$ 0.68   \\
\\
E3a &  14.21 $\pm$ 0.50   &  13.19 $\pm$ 0.48   &  72.60 $\pm$ 0.64   \\ 
E3b &  14.04 $\pm$ 0.50   &  13.57 $\pm$ 0.49   &  72.39 $\pm$ 0.64   \\ 
E3c &  10.63 $\pm$ 0.44   &  12.90 $\pm$ 0.48   &  76.47 $\pm$ 0.61   \\
\\
E4a &  13.05 $\pm$ 0.50   &  10.85 $\pm$ 0.46   &  76.10 $\pm$ 0.63   \\ 
E4b &  12.72 $\pm$ 0.50   &  10.58 $\pm$ 0.46   &  76.70 $\pm$ 0.63   \\ 
E5c &  12.67 $\pm$ 0.49   &  10.11 $\pm$ 0.45   &  77.22 $\pm$ 0.62   \\
 \\
E5a &  21.71 $\pm$ 0.61   &   9.69 $\pm$ 0.44   &  68.60 $\pm$ 0.69   \\ 
E5b &  23.06 $\pm$ 0.63   &   9.38 $\pm$ 0.43   &  67.96 $\pm$ 0.69   \\ 
E5c &  20.92 $\pm$ 0.60   &   8.86 $\pm$ 0.42   &  70.22 $\pm$ 0.68   \\
\hline
\end{tabular} 
\end{table}

\begin{figure}
\vspace{20pt}
\includegraphics[width=84mm]{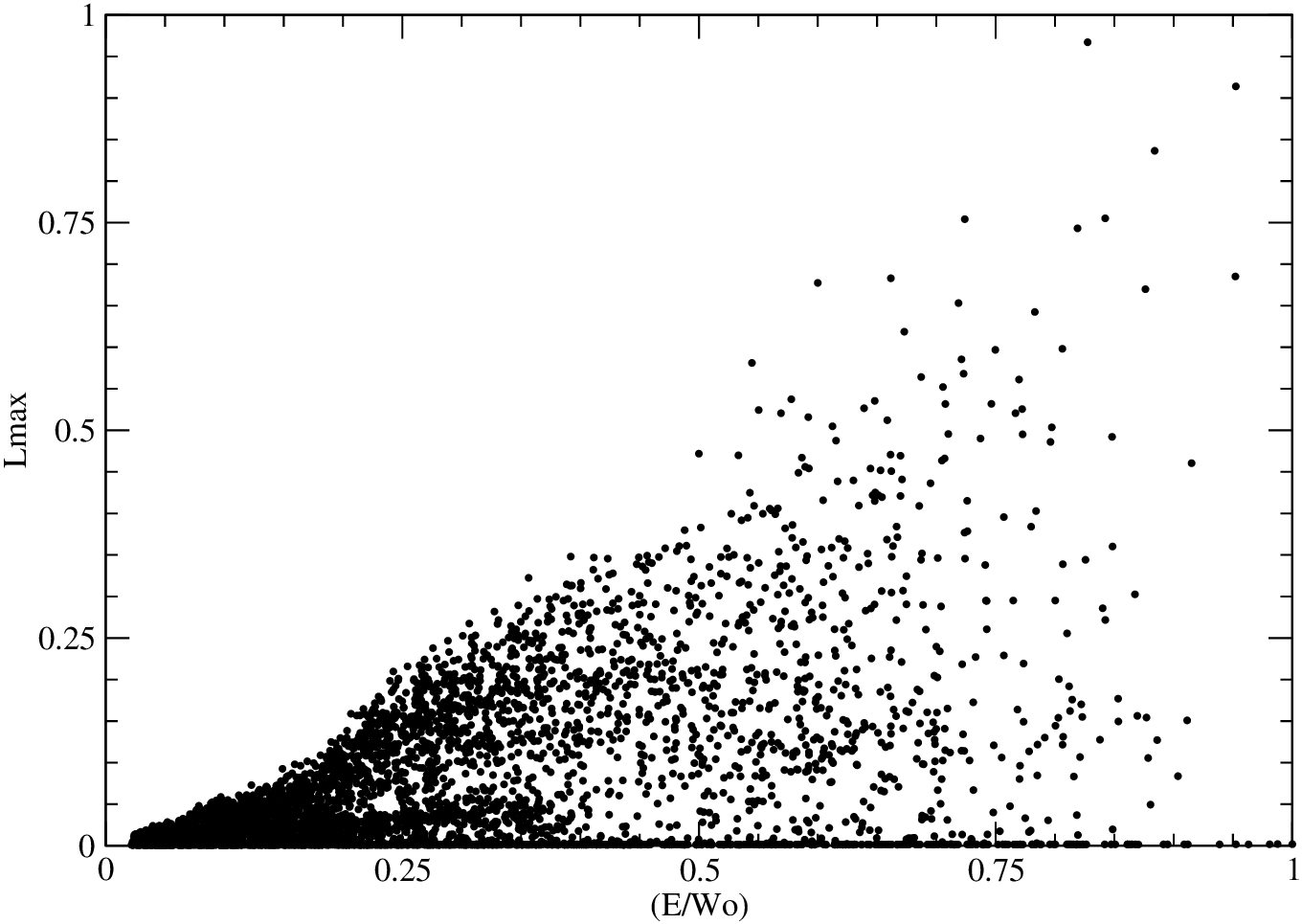}
\caption{The maximum FT-LCNs ($L_{max}$) versus reduced orbital energy plot for Model E5c}
	\label{fig:ervmaxlye5abc}
\end{figure}

The values of the FT-LCNs are larger than those we found for non-cuspy models,
in agreement with the results of \citet{KS2002} and \citet{KS2003}.
Figure \ref{fig:ervmaxlye5abc} shows, for our model E5c, the plot of $L_{max}$ versus 
the reduced energy (i.e., the orbital energy, E, divided by the potential energy
at the galactic centre, $W_{o}$), which can be compared with Fig. 3 of \citet{MCW2005}.
In the present case the largest FT-LCNs are close to $1.0 \:(t.u.)^{-1}$, while for
the non-cuspy model of \citet{MCW2005} they only reached about $0.5 \:(t.u.)^{-1}$.
Besides, in the non-cuspy system there were no chaotic orbits for $E/W_{o}$ values
close to $1.0$ (i.e., orbits very close to the centre of the system),
because the potential near the centre was essentially that of a three dimensional
harmonic oscillator. But, here, the presence of the cusp strongly limits the central
region where the potential can be approximated with an integrable potential and,
therefore, one can find chaotic orbits near the centre of the system.

For each model we took eleven orbital positions at intervals of $10 \:t.u.$,
i.e., over a total interval of $100 \:t.u.$, to obtain the mean square values of each
coordinate separately for each type of orbit. Table~\ref{tab8} gives the $y/x$ and $z/x$
axial ratios, computed from the square roots of those quadratic mean values.
All the results for the a, b and c realizations of a same model show very
good agreement at the $3\sigma$ level, except for the $y/x$ ratio difference
between models E4b and E4c for the fully chaotic orbits, which is somewhat larger.
The differences between the distributions of the
partially and fully chaotic orbits are significant at the $3\sigma$ level
only for $z/x$ ratio of the E5 models and the b case of models E4. This result
agrees with those obtained for the models we studied before \citep{AMNZ2007,
MNZ2009} in the sense that the differences between the distributions of partially
and fully chaotic orbits, on the one hand, increased for
more flattened systems and, on the other hand, diminished for cuspy systems.

\begin{table}
\caption{ Axial ratios of different kinds of orbits.}
\label{tab8}
\begin{tabular}{@{}lcccc@{}}
\hline
$Ratio$ & $Mod$ & $Regular$ & $Part.\; Chaotic$ & $Fully \; Chaotic$ \\ 
\hline 
y/x & E2a &  1.030 $\pm$ 0.015  &  0.903 $\pm$ 0.016  &  0.828 $\pm$ 0.014  \\ 
$ $ & E2b &  1.023 $\pm$ 0.014  &  0.899 $\pm$ 0.016  &  0.841 $\pm$ 0.014  \\ 
$ $ & E2c &  1.003 $\pm$ 0.014  &  0.891 $\pm$ 0.016  &  0.813 $\pm$ 0.014  \\ 
\\  
$ $ & E3a &  0.816 $\pm$ 0.022  &  0.878 $\pm$ 0.020  &  0.800 $\pm$ 0.024  \\ 
$ $ & E3b &  0.815 $\pm$ 0.023  &  0.863 $\pm$ 0.021  &  0.765 $\pm$ 0.027  \\ 
$ $ & E3c &  0.773 $\pm$ 0.025  &  0.915 $\pm$ 0.020  &  0.799 $\pm$ 0.025  \\
\\  
$ $ & E4a &  0.520 $\pm$ 0.040  &  0.804 $\pm$ 0.032  &  0.712 $\pm$ 0.032 \\
$ $ & E4b &  0.510 $\pm$ 0.044  &  0.781 $\pm$ 0.035  &  0.814 $\pm$ 0.034 \\
$ $ & E4c &  0.568 $\pm$ 0.039  &  0.894 $\pm$ 0.030  &  0.652 $\pm$ 0.029 \\
 \\   
$ $ & E5a &  0.892 $\pm$ 0.028  &  0.869 $\pm$ 0.029  &  0.903 $\pm$ 0.018 \\
$ $ & E5b &  0.965 $\pm$ 0.026  &  0.868 $\pm$ 0.030  &  0.871 $\pm$ 0.017 \\
$ $ & E5c &  0.934 $\pm$ 0.029  &  0.790 $\pm$ 0.029  &  0.813 $\pm$ 0.016 \\          
\hline
z/x & E2a &  0.910 $\pm$ 0.016  &  0.905 $\pm$ 0.016  &  0.919 $\pm$ 0.013 \\ 
$ $ & E2b &  0.937 $\pm$ 0.015  &  0.870 $\pm$ 0.016  &  0.917 $\pm$ 0.013 \\ 
$ $ & E2c &  0.879 $\pm$ 0.015  &  0.930 $\pm$ 0.016  &  0.903 $\pm$ 0.014 \\
 \\   
$ $ & E3a &  0.676 $\pm$ 0.024  &  0.825 $\pm$ 0.021  &  0.867 $\pm$ 0.024 \\ 
$ $ & E3b &  0.712 $\pm$ 0.027  &  0.791 $\pm$ 0.021  &  0.831 $\pm$ 0.026 \\ 
$ $ & E3c &  0.668 $\pm$ 0.028  &  0.845 $\pm$ 0.022  &  0.850 $\pm$ 0.025 \\
\\    
$ $ & E4a &  0.424 $\pm$ 0.053  &  0.648 $\pm$ 0.037  &  0.798 $\pm$ 0.034 \\
$ $ & E4b &  0.359 $\pm$ 0.045  &  0.573 $\pm$ 0.037  &  0.789 $\pm$ 0.033 \\
$ $ & E4c &  0.401 $\pm$ 0.049  &  0.699 $\pm$ 0.036  &  0.730 $\pm$ 0.031 \\ 
\\   
$ $ & E5a &  0.309 $\pm$ 0.033  &  0.374 $\pm$ 0.033  & 0.733 $\pm$ 0.020 \\
$ $ & E5b &  0.361 $\pm$ 0.040  &  0.484 $\pm$ 0.045  & 0.675 $\pm$ 0.019 \\
$ $ & E5c &  0.319 $\pm$ 0.043  &  0.398 $\pm$ 0.038  & 0.671 $\pm$ 0.018 \\ 
\hline
\end{tabular} 
\end{table} 

\section{Conclusions}

The method of \citet{HO1992} is clearly better than that of Aguilar
\citep{AM1990} to build cuspy triaxial stellar models, since the former needs
no softening and uses a radial expansion which is particularly adequate to fit
N--body distributions with $\gamma \simeq 1$. As a side benefit, the coefficients
of both the radial and angular expansions are provided by the method itself and
there is no need of additional fittings to obtain them. The only caveat is
that, as shown by \citet{KEV2008}, the method of Hernquist and Ostriker is
adequate for cases with $\gamma \simeq 1$ only and, for other slopes of the cusp,
other related but different methods should be preferred.

The models obtained using the same parameters and changing only the seed number
of the random number generator show that our results are very robust, indeed. Those
models exhibit essentially the same global properties (except for the angular velocity
of figure rotation) and, with very few exceptions, the same fractions and distributions
of regular, partially and fully chaotic orbits. The different angular velocities we
obtained here for statistically equivalent models further complicates the little we
know about the phenomenon of figure rotation (i.e., with zero angular momentum) in
this kind of models. On the one hand, the reality of the rotation has been checked
using both the Aguilar \citep{AM1990} and Aarseth \citep{A2003} codes by \citet{M2006} and here
with the Hernquist code and, on the other hand, such triaxial systems with figure rotation are
just the stellar equivalent of the Riemann ellipsoids of fluid dynamics, so that there is
nothing misterious about their existence. Nevertheless, except for the increase of the
angular velocity when one goes to flatter systems, the rotation does not seem to correlate
with other properties of the systems and, moreover, now it turns out that statistically
equivalent systems have different velocities. Even though, as already indicated, the
angular velocities are so low that they have very little effect on orbital chaoticity,
this phenomenon is interesting and warrants further investigation.

The high stability of our models over time scales of the order of a Hubble time is very
well stablished. Not only are the detected variations very small, but they are most likely
due to relaxation effects of the N--body code. Besides, it should be stressed that we have
checked the stability running the models self-consistently, while checks on the
stability of models resulting from the method of \citet{S1979} are usually done on
fixed potentials \citep{S1993}. As we have shown, when the potential is fixed, the changes
in our models are an order of magnitude smaller than those found self-consistently.

Although, considering the results of \citet{KS2002}, \citet{KS2003} and \citet{MNZ2009}, high
fractions of chaotic orbits could be expected in our cuspy triaxial models, the fractions found
are extremely high, indeed, exceeding $85$ per cent for our E2 and E3 models. It is important
to emphasize that those high fractions of chaotic orbits are not merely due to the
low $L_{lim}$ adopted: Had we adopted the $0.010 \:(t.u.)^{-1}$ limit, those models would
still have had between $70$ an $75$ per cent of chaotic orbits.

The main conclusion of our investigation is that it is perfectly possible to build
self-consistent stellar models of cuspy triaxial elliptical galaxies that are very stable 
despite containing high percentages of chaotic orbits. This result confirms and extends
previous work by others \citep{VKS2002, KV2005} and ours \citep{AMNZ2007, MNZ2009}. It also
supports the suggestion by \citet{MCW2005} in the  sense that the difficulties to obtain
such models with the method of Schwarzschild should be attributed to the method itself and
not to physical causes. It is true that, while regular orbits always occupy the same region
of space, chaotic orbits may spend a long interval occupying certain region, only to switch
to a different one later on, with sticky orbits being an extreme example. But, although such
behaviour conspires against obtaining stable models with high fractions of chaotic orbits
with the method of Schwarzschild, it does not necessarily prevents the existence of those
models. All one needs is that, as some chaotic orbits abandon a region of space to explore
a different one, orbits in the latter region move on to replace those in the former. In
other words, rather than having a static equilibrium, with each orbit covering always the
same zone, we have a dynamic equilibrium where switches from one zone to another are
present but balance each other on average. Our models show that such state of affairs is
perfectly possible.

\section{Acknowledgements}
We are very grateful to Lars Hernquist and Daniel Pfenniger for allowing us to
use their codes, to C. Efthymiopoulos for useful comments, to H.D. Navone for his
assistance with programming, to R.E. Mart\'{\i}nez and H.R Viturro
for their  technical assistance and to an anonymous referee for his comments on
the original version of the present paper. This work was supported with grants
from the Consejo Nacional de Investigaciones Cient\'{\i}ficas y T\'ecnicas de la
Rep\'ublica Argentina, the Agencia Nacional de Promoci\'on Cient\'{\i}fica y
Tecnol\'ogica, The Universidad Nacional de La Plata and the Universidad Nacional de
Rosario.

\end{document}